\documentclass[aps,prx,notitlepage,superscriptaddress,showpacs,twocolumn ]{revtex4-1}
\usepackage{graphicx,subfigure,epsfig}
 \usepackage{array,multirow}
\usepackage{dcolumn}
\usepackage{amssymb,amsmath,amsfonts,mathrsfs}
\usepackage{array}
\usepackage{times,setspace}
\usepackage{latexsym}
\usepackage{float,flafter,bm,bbm}
\usepackage{epstopdf,color,multirow}
\usepackage[colorlinks,linkcolor=blue,anchorcolor=blue,urlcolor=blue,citecolor=blue]{hyperref}
\usepackage{footnote}
 \usepackage[OT1]{fontenc}
\usepackage[normalem]{ulem}
\usepackage{xcolor}

\hypersetup{
    colorlinks=true,
    linkcolor=blue,
    filecolor=magenta,
    urlcolor=blue,
}

\newcommand{\ket}[1]{\left\vert #1 \right\rangle}

\newcommand{\ketbra}[2]{\left| #1\right>\left< #2 \right|}
\newcommand{\braket}[2]{\left< #1| #2 \right>}

\begin{document}

\title{Inverting multiple quantum many-body scars via disorder}

\author{Qianqian Chen}
\affiliation{Kavli Institute for Theoretical Sciences, University of Chinese Academy
of Sciences, Beijing 100190, China}
\author{Zheng Zhu}
\email{zhuzheng@ucas.ac.cn}
\affiliation{Kavli Institute for Theoretical Sciences, University of Chinese Academy of Sciences, Beijing 100190, China}
\affiliation{CAS Center for Excellence in Topological Quantum Computation, University of Chinese Academy of Sciences, Beijing, 100190, China}

\begin{abstract}
The observations of persistent revivals in the Rydberg atom chain have revealed a weak ergodicity breaking mechanism known as quantum many-body scars, which is typically a collection of  states with low entanglement embedded in otherwise thermal spectra. Here, by applying a generic formalism, we reveal a direct evolution from the quantum many-body scars to the multiple inverted quantum many-body scars, i.e., different sets of excited states with volume-law entanglement entropy embedded in a sea of states with area-law entanglement. When increasing the disorder strength, a series of exact eigenstates, acting as conventional QMBS in a regime of weak disorder, remain unchanged. Around each of these states, the inverted quantum many-body scars are introduced by the increased disorder. Moreover, the strong disorder also gives rise to additional sets of inverted quantum many-body scars with their energies concentrating in the middle of the exact eigenstates. As a result, all the multiple inverted quantum many-body scars are approximately equidistant in energy, reminiscent of conventional quantum many-body scarred states. Despite being the measure-zero states in the whole spectrum, these inverted quantum many-body scarred states significantly influence nonequilibrium dynamics in the large disorder regime. Random thermal states in a specific subspace show periodic revivals in the fidelity dynamics, while the typical charge-density-wave states exhibit persistent imbalance dynamics. We further examine the stability of the conventional and the inverted quantum many-body scars against the external random field. Our findings expand the variety of nonthermal systems and draw a connection between the weak violation of ergodicity and that of non-ergodicity.
\end{abstract}

\date{\today}
\maketitle

\section {Introduction}
Most isolated quantum many-body systems evolve into an equilibrium statistical ensemble under the mechanism of quantum ergodicity~\cite{Srednicki1994,Dunjko2008Nature, Nandkishore2015,DAlessio2016}.
Due to the quest to realize long-lived coherent dynamics, tremendous attempts have been made to develop ergodicity-breaking mechanisms.
Among the very few exceptions of quantum ergodicity in isolated systems, a weak ergodicity-breaking system with the so-called quantum many-body scar (QMBS) states~\cite{Turner2018NP, Turner2018PRB,Papic2021Nature, Papic2021arXiv, Nicolas2021,Chandran2022}  has recently garnered intense interest, and was realized in ultracold-atom experiments~\cite{Bernien2017,Bluvstein2021, GuoXianSu2022}.
Having only a few conserved quantities and being typically disorder-free, QMBS is characterized by certain initial states that periodically revive and is comprised of isolated nonthermal eigenstates embedded in a sea of thermal states.
These features are significantly distinguished from the previously known strong ergodicity-breaking mechanisms, i.e., integrable systems with an extensive number of conserved quantities~\cite{Deutsch1991PRA,Kinoshita2006Nature, 2009PhRvL103j0403R, 2010PhRvL105y0401B} and many-body localization  (MBL)~\cite{Basko2006,Nandkishore2015,Altman2015,DongLingDeng2017, Parameswaran2018, Abanin2019,Abanin2021} with low entangled eigenstates  in the presence of strong disorder typically. Moreover, beyond isolated systems, QMBS states have also been  found in the relevant contexts of open quantum  systems \cite{Buca2019,Pakrouski2021,QianqianChen2022}.

Recently, the exploration of QMBS has taken an intriguing turn with efforts aimed at conceptualizing their inverse. 
This refers to highly entangled  excited states with volume-law entanglement embedded in a sea of states with area-law entanglement.
This phenomenon, known as inverted QMBS, diversifies the landscape of nonthermal quantum systems. Previous studies \cite{Srivatsa2020,Iversen2022,Srivatsa2022} in this regard focused on non-thermal states in a single narrow energy window. However, a critical aspect that remains uncertain is the potential for inverted QMBS to manifest in multiple energy windows, especially with energies that are approximately or precisely equidistant. 
Additionally, unlike the  unified formalisms of QMBS  \cite{Shiraishi2017,Schecter2019,Chattopadhyay2020, Mark2020AKLT, Moudgalya2020etaPairing, Mark2020Hubbard, Medenjak2020,ODea2020,Pakrouski2020,JieRen2021, Moudgalya2022CommutantAlgebra}, the systematic formalism to construct the inverted QMBS that resembles thermal states is still elusive. Addressing this gap could offer profound insights into the intricate dynamics of quantum systems and potentially lead to novel applications in quantum computing and information science.

On the other hand, the connections between distinct ergodicity-breaking mechanisms lie at the core of understanding thermalization and its absence. Indeed, the disorder, which is ubiquitous across the realistic quantum simulators \cite{Smith2016,JiehangZhang2017,Marcuzzi2017}, can  bring integrability, MBL, and QMBS together \cite{ChunChen2018,Shibata2020,Moudgalya2020etaPairing,MondragonShem2021, KeHuang2021,Voorden2021, LianBiao2022,Halimeh2022, Tamura2022,GZhang2022}. According to recent studies \cite{MondragonShem2021,KeHuang2021} of QMBS in PXP models, in the process of increasing the disorder,
the system is always first deprived of the original QMBS and becomes fully thermal, and then the possible transition/crossover to MBL emerges.
The mechanisms causing scars in the PXP model  are only approximately understood \cite{Khemani2019Apr,Choi2019, WenWeiHo2019, ChengJuLin2019, Iadecola2019, Bull2020, Michailidis2020PRX, Surace2020,  Alhambra2020, Magnifico2020, Desaules2022, *Desaules2022truncatedSchwinger},
then it is  fundamentally important to explore the exact QMBS that is analytically tractable \cite{Moudgalya2018nonintegrable,Moudgalya2018AKLTEE,Schecter2019,Ok2019,Iadecola2020,Mark2020AKLT, Chattopadhyay2020,Moudgalya2020etaPairing, Moudgalya2020MPS, Mark2020Hubbard, Langlett2022} in the presence of the disorder with the interplay of different ergodicity-breaking mechanisms. In particular, the direct evolution from a system with exact QMBS to the one with an MBL background has not been revealed.

Since both conventional and inverted QMBS are a small fraction of states that have very different thermalization properties from other excited states, here we realize them under a same formalism and invert them directly through disorder. We study a typical disordered model that represents a large class of Hamiltonians with a tower of exact QMBS states at weak disorder. Then we increase the disorder strength and drive the majority of the states to be many-body localized, while the original exact QMBS eigenstates are protected as invariable in the whole process. 
Moreover, the large disorder also introduce multiple inverted QMBS states at finite energy density within the MBL spectrum, as characterized by a collection of states with volume-law entanglement entropy (EE) while the whole spectrum follows Poisson statistics. 
We further explore the nature and the properties of the inverted QMBS. We discover that the inverted QMBS are states superposed with randomness by the null vectors of the disorder. Despite such fundamentally different physical implications of inverted QMBS compared to their conventional counterparts, both share certain conceptual similarities. Firstly, both inverted and conventional QMBS comprise sets of states that are equidistant in energy. Secondly, both are considered measure-zero states in their respective spectrums. Thirdly, they can manifest periodic revivals in fidelity dynamics; however, for inverted QMBS, such dynamics emerge from a random thermal state within a specific  subspace.
We also apply the onsite random field to examine the stability of the scarring states, and find both the original exact QMBS and the inverted QMBS states disappear with increasing onsite randomness.

\section {Generic  framework for inverted QMBS}

We consider a generic framework for QMBS \cite{ODea2020,Pakrouski2020,JieRen2021} and also apply it to realize the inverted QMBS.
Such a framework constructs a Hamiltonian
\begin{equation}\label{eq:framework}
H=H_{\mathrm{sym}}+H_{\mathrm{SG}}+H_\text{A}.
\end{equation}
Here, $H_\mathrm{sym}$ is $G$-symmetric with $[H_{\mathrm{sym}}, Q^{+}]=0$ and $[H_{\mathrm{sym}}, H_\mathrm{SG}]=0$, where $G$ is a non-Abelian symmetry and $Q^+$ is the  spectrum-generating ``ladder" operator.
The second term $H_{\mathrm{SG}}$ is a linear combination of generators in the Cartan subalgebra of $G$, and fulfills a spectrum-generating algebra (SGA)
$
 [H_{\mathrm{SG}}, Q^{+}]=\omega Q^{+}
$
that can lead to a tower of eigenstates $\ket{\mathcal{S}_k}$ with energy  spacing $\omega$ and low EE.
Here, $\{\ket{\mathcal{S}_k}\}_k$ is a particular set of eigenstates and labeled by the eigenvalue under the Casimir operators of $G$, and states in the set are distinguished by their eigenvalues under Cartan generators of $G$.
The term $H_\mathrm{A}$ breaks the $G$-symmetry, and   is  immaterial to the dynamics of the scarred eigenstates $\{\ket{\mathcal{S}_k}\}_k$ since it annihilates them $H_\mathrm{A}\ket{\mathcal{S}_k}=0$.
By noting that scarred states $\ket{\mathcal{S}_k}$ distinguish itself by being the superposition of the null vector of $H_\mathrm{A}$, which is a fundamental element for determining the thermalization properties of the total Hamiltonian, we generalize the framework to construct the multiple inverted  QMBS states as random superpositions. Specifically, when considering a disordered version of $H_\mathrm{A}$ with a large amplitude, considerable null vectors, and the potential for inducing low entanglement for majority states, we can effectively construct the multiple inverted QMBS.
In the following, we will apply this framework to show an exemplary case that realizes both conventional and multiple inverted  QMBS states.

We adopt $H_\mathrm{sym}$ as the $S=1/2$ XX Heisenberg chain that is potentially realizable in Rydberg quantum simulators \cite{Marcuzzi2017,Ostmann2019,VanVoorden2021,XXLi2022}
\begin{equation}\label{eq:HSym}
  H_\mathrm{sym}=\sum_{j=1}^{L} S_{j}^{+} S_{j+1}^{-}+S_{j}^{-} S_{j+1}^{+}.
\end{equation}
$H_\mathrm{sym}$ is integral
\cite{Franchini2017} and has the
Onsager symmetry \cite{Vernier2019,Shibata2020}, i.e., $H_\mathrm{sym}$ commutes with all the Onsager-algebra elements, including
\begin{equation}\label{}
  Q=\sum_{j=1}^{L} S_{j}^{z}, \quad Q^{+}=\sum_{j=1}^L(-1)^{j+1} S_{j}^{+} S_{j+1}^{+}.
\end{equation}
Letting $H_\mathrm{SG}=Q$, we have the SGA
\begin{equation}\label{eq:SGA}
  [H_\mathrm{SG},Q^+]=2 Q^+.
\end{equation}
Due to \eqref{eq:SGA}, the set of degenerate states
\begin{equation}\label{eq:scar}
  \ket{\mathcal{S}_k}=(Q^+)^k\ket{\Downarrow}\quad (k=0,\dots,\lfloor L/2\rfloor)
\end{equation}
of $H_\mathrm{sym}$ can be lifted and promoted to the evenly spaced exact tower of eigenstates with energies $E_\mathcal{S}=-L/2+2k$. Here, $\ket{\Downarrow}$ denotes a polarized spin-down state. Finally, $H_{\mathrm{A}}$ is added to destroy the integrability  and annihilate each of the $\{\ket{\mathcal{S}_k}\}_k$.
Furthermore, we choose a disordered term $H_{\mathrm{A}}$ which can drive the majority states to be MBL when increasing the disorder strength,
\begin{equation}\label{eq:HAc1}
  H_{\mathrm{A}}=\Delta\sum_{j=1}^L\left\{c _ { j }  | 0 1 0 \rangle \langle010|\right\}_{j-1, j, j+1},
\end{equation}
where $c _ { j }$   are  the uniform random numbers $c _ { j } \in[-1,1]$, and $\Delta$ denotes the disorder strength.
In the following, we will show that $H_{\mathrm{A}}$ preserves not only the exact special states $\{\ket{\mathcal{S}_k}\}_k$ but also a set of states with higher EE than the MBL states.
To summarize, the total Hamiltonian reads
\begin{equation}\label{eq:totH}
\begin{aligned}
  H=&\sum_{j=1}^{L} \left(S_{j}^{+} S_{j+1}^{-}+S_{j}^{-} S_{j+1}^{+}+ h_0S_{j}^{z}\right)\\
  &+\Delta\sum_{j=1}^{L}\left\{c _ { j }  | 0 1 0 \rangle \langle010|\right\}_{j-1, j, j+1}.
\end{aligned}
\end{equation}
Unless otherwise stated, we choose $h_0=1$.
We use the exact diagonalization (ED) approach to examine the whole spectrum of the model~\eqref{eq:totH}. In the following, we mainly focus on the bulk $S_\mathrm{tot}^z$ sectors, and we average data over $10\sim100$ disorder realizations (denoted as $[\cdot]$) depending on the system size $L$ and the $S_\mathrm{tot}^z$ sectors.

\begin{figure}[tbp]
\begin{center}
\includegraphics[width=0.5\textwidth]{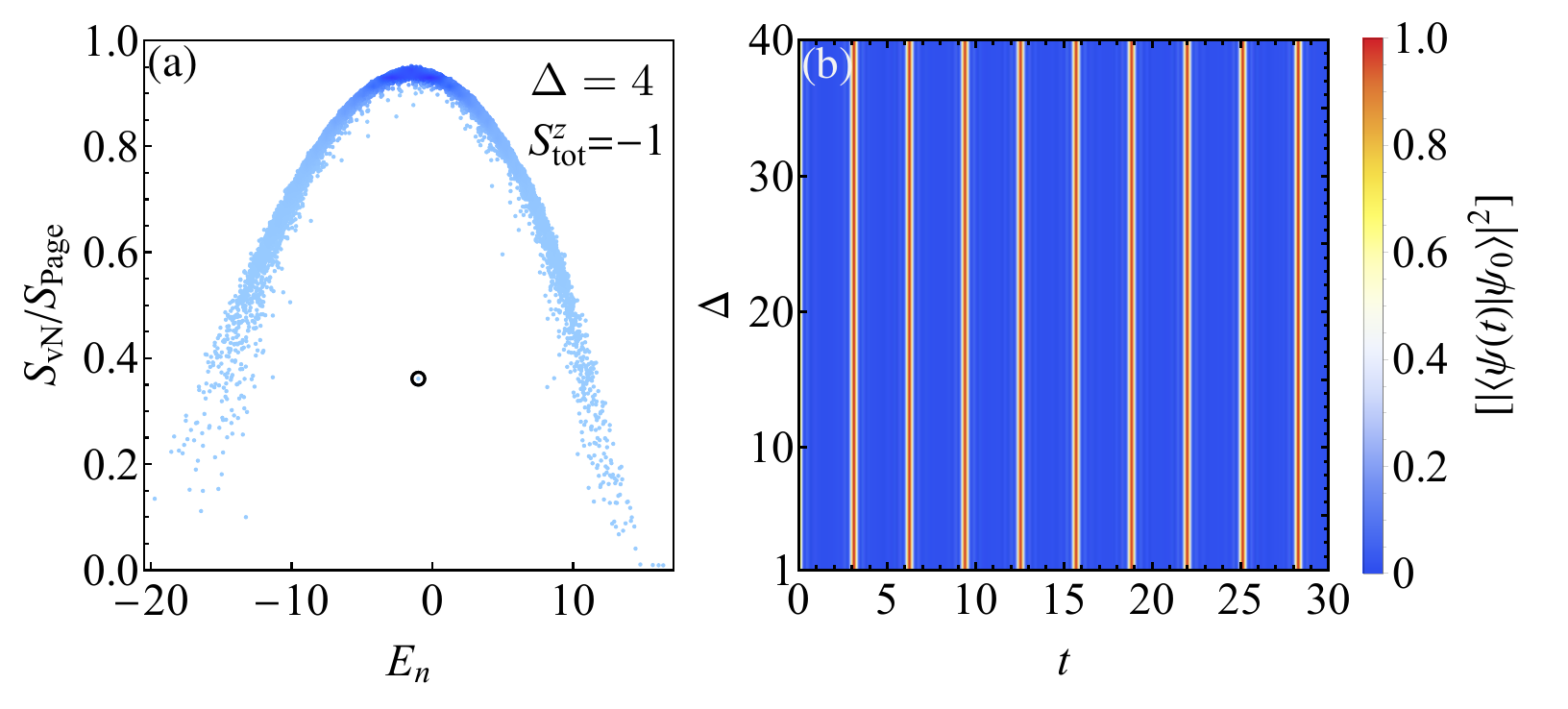}
\end{center}
\par
\renewcommand{\figurename}{Fig.}
\caption{
Typical features of exact quantum many-body scar.
(a) $S_\text{vN}/S_\text{Page}$ with respect to all eigenenergies at weak disorder for $L=18$.
The black circle denotes the scarred state $\ket{\mathcal{S}_4}$.
Darker colors imply a higher density of the states.
(b) The disorder-averaged fidelity dynamics $[|\braket{\psi(t)}{\psi(0)}|^{2}]$ of the initial  state $\ket{\psi(0)}\equiv\ket{\psi_0}$ in scar subspace as a function of disorder strength $\Delta$ when $L=18$.
 }
\label{Fig1_scar}
\end{figure}
\begin{figure}[tbp]
\begin{center}
\includegraphics[width=0.5\textwidth]{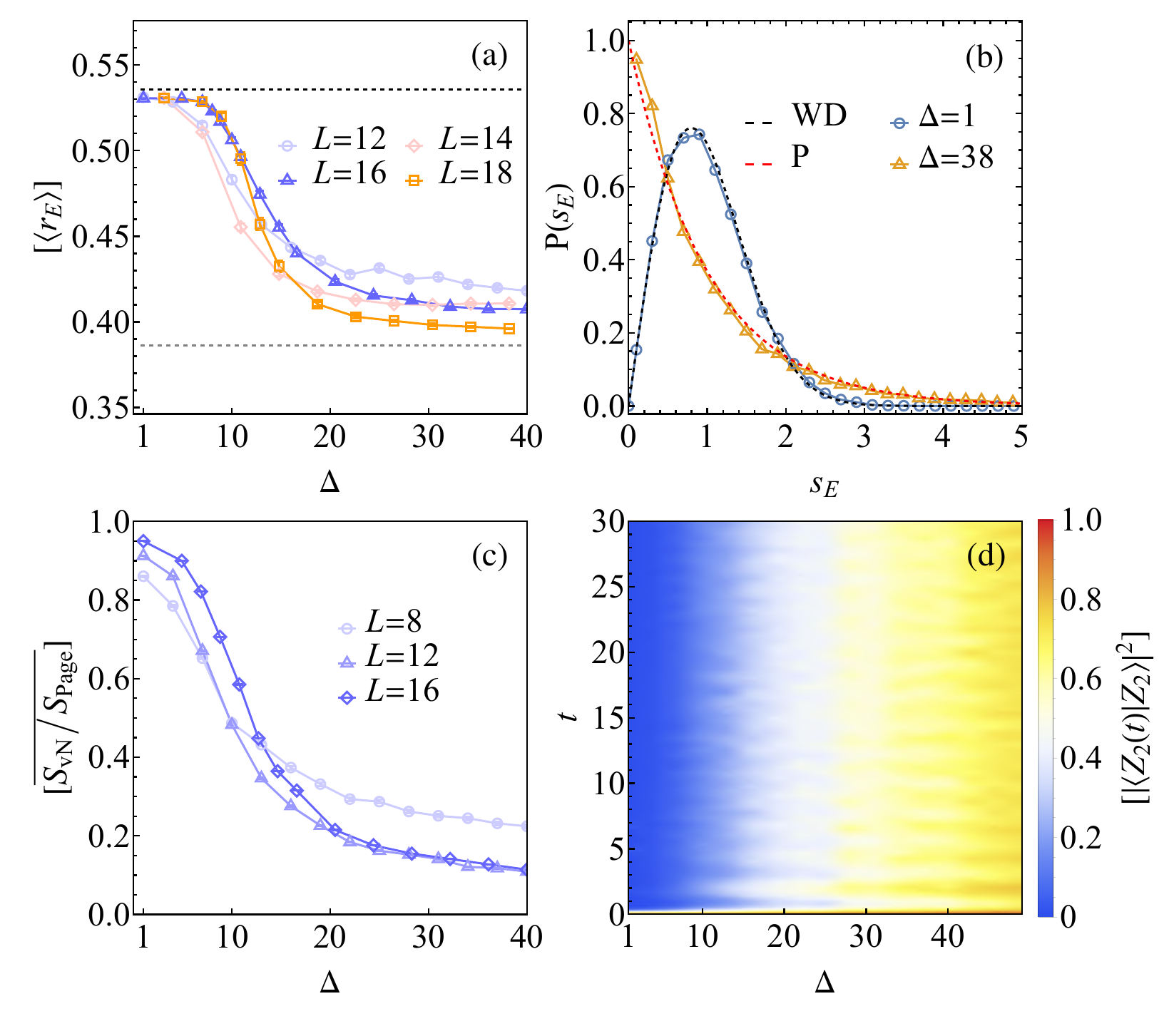}
\end{center}
\par
\renewcommand{\figurename}{Fig.}
\caption{
The nature of  bulk states as functions of disorder strength $\Delta$.
(a) Mean level spacing ratios $[\langle r_E\rangle]$ for eigenenergies in the middle 60\% of the spectrum with $S^z_\text{tot}=0$ for $L=12,16$ and $S^z_\text{tot}=-1$ for $L=14,18$. As a comparison, Wigner-Dyson (WD) statistics of the GOE $\left\langle r_E\right\rangle\approx 0.536$ (dashed black lines) and Poisson (P) statistics  $\left\langle r_E\right\rangle\approx 0.38$ (dashed gray lines) are plotted.
(b) The energy level spacing statistics for one particular disorder realization in $S^z=-1$ sector with $L=18$, after  performing the spectrum unfolding.
(c) $[\overline{S_\text{vN}/S_\text{Page}}]$ as a function of $\Delta$ with $S^z_\text{tot}=0$.
The data are averaged over $100$ disorder realizations and  over $1/2$ (but not $1/12$) of all the eigenstates that are around the state $\ket{\mathcal{S}_4}$.
(d) The disorder-averaged fidelity dynamics $[|\braket{Z_2(t)}{Z_2}|^{2}]$ of the initial  state $\ket{Z_2}$ with $L=14$ at different disorder strength $\Delta$.
}
\label{Fig:Bulk}
\end{figure}

\section {conventional QMBS }
The exact tower of eigenstates $\ket{\mathcal{S}_k}$  exhibit exactly equal energy spacing and persevere at any disorder strength $\Delta$. They are conventional exact QMBS states embedded in otherwise thermal spectra at smaller $\Delta$,
while at larger $\Delta$, they are embedded in MBL spectra. Below we reveal their nature from the eigenstate EE and the fidelity dynamics.

A wealth of thermalization information on physical states can be obtained from the EE. We consider the density matrix \(\rho_{n}\) of the \(n\)th eigenstate $\ket{\phi_n}$ defined
by $\rho_{n} = \left| \phi_{n} \right\rangle\left\langle \phi_{n} \right|$, and study the EE
$
    S_{\text{vN}} = - \text{Tr}_{A}\left( \rho_{{A},n}\ln \rho_{{A},n} \right),
$
where \(\rho_{A,n}\) is the reduced density matrix for subsystem \(A\) (chosen as half chain here) after tracing out the rest of the system.
Figure~\ref{Fig1_scar}(a) shows one typical example of EE at small $\Delta$ for $S_\mathrm{tot}^z=-1$.
The majority of the bulk eigenstates have EE approaching the Page value for a random pure state \cite{Page1993} $S_\mathrm{Page}\approx\ln (\mathcal{D}_A)-0.5{\mathcal{D}_A}/{\mathcal{D}_B}$, where $\mathcal{D}_A$ ($\mathcal{D}_B$) is the Hilbert space dimensions of subsystem $A$ ($B$),  while the scarred state $\ket{\mathcal{S}_4}$ (marked by a black circle) exhibits anomaly low EE. $S_\mathrm{tot}^z$ sectors  with other eigenstates $\ket{\mathcal{S}_k}$ exhibit similar behavior at small disorder strength. In the following, we will show that, with the increase of the disorder strength, the energy-level statistics of the bulk energy spectra change from Wigner-Dyson to  Poisson, while the EE of the exact tower of states $\ket{\mathcal{S}_k}$ remains the same for any disorder strength $\Delta$.

The existence of exact QMBS can also be inferred by the
fidelity dynamics for specific initial states in the scar subspace. Figure~\ref{Fig1_scar}(b) shows  perfectly periodic revivals in the fidelity of the initial state $\ket{\psi_0} =\frac{1}{\mathcal{N}}\sum_{k=0}^{\lfloor L/2\rfloor}\ket{\mathcal{S}_k}$, where $\mathcal{N}$ is the normalization factor. The revival period \(T=\pi\) corresponds to the energy interval \(\omega=2\)  of the scarred states $\ket{\mathcal{S}_k}$, as expected from the SGA \eqref{eq:SGA}. Since the disorder term $H_\mathrm{A}$ annihilates $\ket{\mathcal{S}_k}$,
the exact states $\ket{\mathcal{S}_k}$ and the  fidelity dynamics are preserved regardless of the disorder strength $\Delta$, as Fig.~\ref{Fig1_scar}(b) shows.

\section {Ergodic to non-ergodic background} Although increasing the disorder strength in $H_\mathrm{A}$ does not influence the eigenstates $\ket{\mathcal{S}_k}$, most of the other bulk states alter from ergodic to non-ergodic.
To show this, we examine the energy level statistics, half-chain EE, and the imbalance dynamics as a function of disorder strength $\Delta$, as shown in Figs.~\ref{Fig:Bulk}. Here we remark that the existence and stability of MBL in thermodynamics is under active debate \cite{Roeck2017,Suntajs2020,Sels2021,Abanin2021,Crowley2022}. Although our following finite-size numerics do not allow us to infer the existence and stability of MBL in the thermodynamic limit, the realm of finite systems is nonetheless intriguing and pertinent in and of itself, for example, for contemporary experiments using platforms like cold atoms, superconducting processors, and trapped ions.

The  energy-level spacing ratios are defined by \cite{Oganesyan2007}
$
{r}_{E}=({\min (s_{E_n}, s_{E_{n-1}})})/({\max (s_{E_n}, s_{E_{n-1}})}),
$
where
$
  s_{E_n} = E_{n+1} -E_{n}
$
is the nearest-neighbor energy-level spacings and $E_{n}$ is an increasing-ordered set of energy levels. We eliminate $20 \%$ of the eigenenergies at the spectrum's edges when calculating the statistics of energy-level spacings.
The mean energy-level spacing ratios $[\langle r_E\rangle]$ as  functions of $\Delta$ are depicted in Fig.~\ref{Fig:Bulk}(a), with $\langle\cdot\rangle$ denoting the average over the spectrum.
In the Fig.~\ref{Fig:Bulk}(a), we find  that $[\langle r_E\rangle]$ converges to the value of Wigner-Dyson statistics of the Gaussian orthogonal ensemble (GOE) when $\Delta$ is small, implying the thermalization of the bulk states, and $[\langle r_E\rangle]$ approaches to the value of Poisson statistics at larger $\Delta$,  indicating that the system is localized \cite{Atas2013}. For a single disorder realization in each of these two different regimes, typical profiles of the energy-level spacing distributions in Fig.~\ref{Fig:Bulk}(b) are consistent with the disorder-averaged energy-level spacing ratios $[\langle r_E\rangle]$.

We then look at the characteristics of the state- and disorder-averaged EE $S_\mathrm{vN}$ that distinguishes thermalization from MBL \cite{Khemani2017PRL}. Figure~\ref{Fig:Bulk}(c) show  divided by $S_\mathrm{Page}$ for various system size $L$ in the $S^z_\text{tot}=0$ sector, where we denote the state average as $\bar{\cdot}$.  With increasing disorder strength $\Delta$, $[\overline{S_\text{vN}/S_\text{Page}}]$ goes from 1 of the thermalized states to 0 of the many-body localized states.

\begin{figure}[tbp]
\begin{center}
\includegraphics[width=0.5\textwidth]{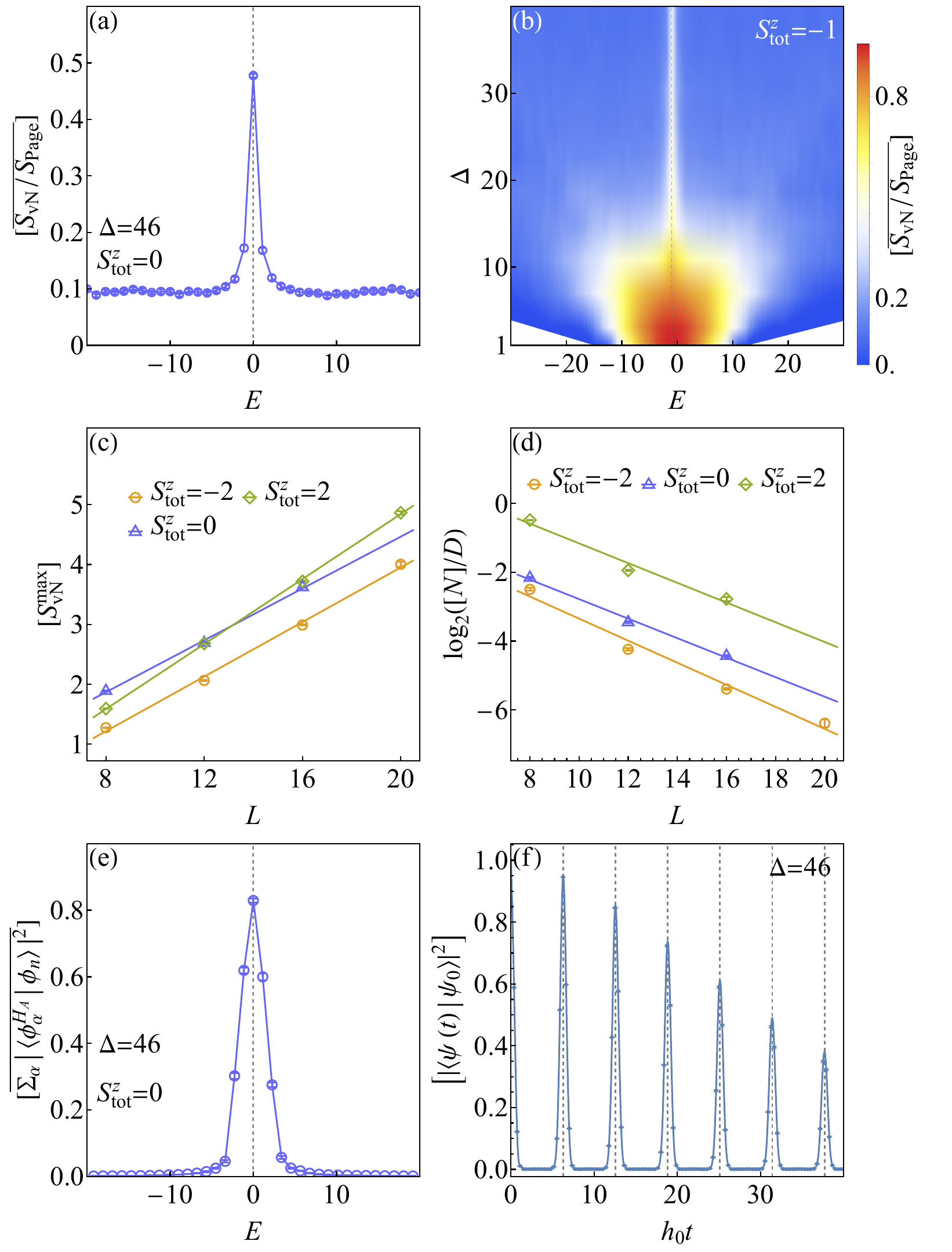}
\end{center}
\par
\renewcommand{\figurename}{Fig.}
\caption{ Energy-resolved features of  states in   $S_\mathrm{tot}^z$ sectors.
(a) The energy-resolved $[\overline{S_\text{vN}/S_\text{Page}}]$  for $L=16$ in $S^z_\text{tot}=0$ sector.
The peak appears at the energy window that has 722 eigenenergies on average.
(b) The energy-resolved $[\overline{S_\text{vN}/S_\text{Page}}]$ as a function of $\Delta$ for $L=18$, $S^z_\text{tot}=-1$.
The bright vertical line resides in the energy window of $E=-1$.
(c) The scaling of $[S_\text{vN}^\text{max}]$ with $L$, where  $[S_\text{vN}^\text{max}]$  are averaged over the maximum entropies  $S_\text{vN}^\text{max}$ of eigenstates in each disorder realization, with the corresponding averaged energies very close to $E_\mathcal{S}$.
(d) The scaling of $\log_2([N]/D)$ with $L$, where $N$ is the number of states with $S_\mathrm{vN}> 0.4S_\mathrm{Page}$, and $D$ is the Hilbert space dimension of $S^z_\mathrm{tot}$.
(e) The overlap between eigenstates of $H$ (i.e., $\ket{\phi_n}$) and   $\vert\phi_\alpha^{H_\mathrm{A}}\rangle$ for $L=16$, where $H_\mathrm{A}\vert\phi_\alpha^{H_\mathrm{A}}\rangle=0$.
{(f) The disorder-averaged fidelity dynamics $[|\langle\psi(t) \mid \psi_0\rangle|^2]$ with $L=14$, where $|\psi_0\rangle$ is a random thermal state in the null space of $H_\mathrm{A}$. Gray vertical dashed lines are plotted with separation $2 \pi$, manifesting that the period of the revival in fidelity dynamics is $2 \pi / h_0$.}
}
\label{Fig:EnergyResolved}
\end{figure}

We also choose the initial product state \(\left| {Z}_{2}\rangle \equiv \right|101010\ldots\rangle\) to represent the imbalance and calculate its dynamics at different $\Delta$. As shown in Fig.~\ref{Fig:Bulk}(d), the fidelity rapidly approaches zero  as time evolves at small $\Delta$, demonstrating ergodic behavior. In contrast, at larger $\Delta$, the persistent imbalance dynamics indicate a non-ergodic evolution, consistent with MBL.

\begin{figure}[tbp]
\begin{center}
\includegraphics[width=0.5\textwidth]{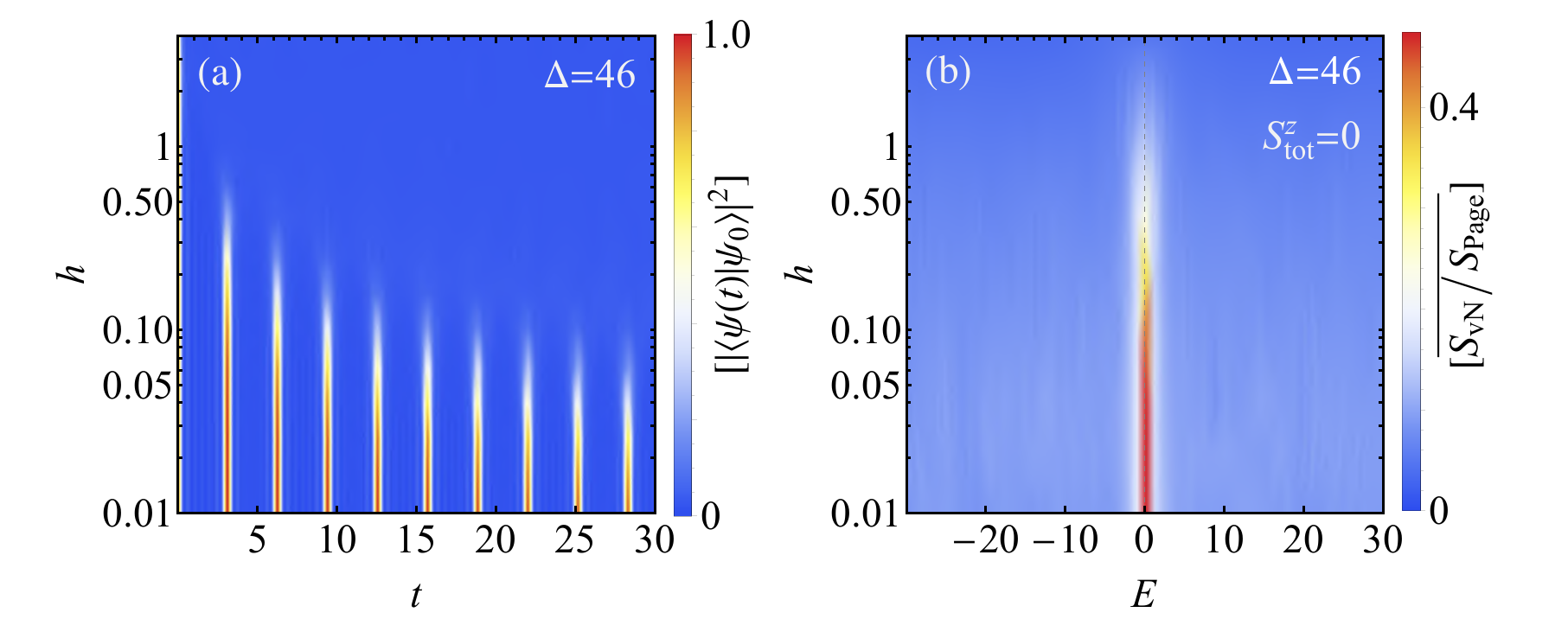}
\end{center}
\par
\renewcommand{\figurename}{Fig.}
\caption{Stability of $\ket{\mathcal{S}_k}$ and the inverted QMBS at large disorder strength $\Delta$. (a) The disorder-averaged fidelity dynamics $[|\braket{\psi(t)}{\psi(0)}|^{2}]$
of the initial  state $\ket{\psi(0)}\equiv\ket{\psi_0}$ in scar subspace with $L=14$.
(b)  The energy-resolved $[\overline{S_\text{vN}/S_\text{Page}}]$ as a function of $h$ for $L=16$, $S^z_\text{tot}=0$.
The   peak resides in the energy window that includes $E_\mathcal{S}=0$ of  the state $\ket{\mathcal{S}_4}$.
}
\label{Fig:c1h}
\end{figure}

\section {Multiple inverted QMBS} 
We have demonstrated that the majority of the bulk states change from ergodic to non-ergodic when increasing disorder strength, below we will show the inverted QMBS states with anomaly high entanglement in the MBL background.

We examine the energy-resolved EE $[\overline{S_\text{vN}/S_\text{Page}}]$ that is averaged over disorder realizations and states in the targeted $S_\mathrm{tot}^z$ sector.
We first look into the $S_\mathrm{tot}^z$ sectors with states $\ket{\mathcal{S}_k}$.
As illustrated in Figs.~\ref{Fig:EnergyResolved}(a,b), we find highly entangled states located very close to $\ket{\mathcal{S}_k}$ in energy,
in sharp contrast to other localized states with low entanglement.
For instance, in Fig.~\ref{Fig:EnergyResolved}(a), the state $\ket{\mathcal{S}_4}$ residing in the sector $S_\mathrm{tot}^z=0$ for $L=16$ has its energy $E_\mathcal{S}=0$, besides which many  highly entangled states jointly give a $[\overline{S_\text{vN}/S_\text{Page}}]$ peak at the energy window of $E=0$.
Figure~\ref{Fig:EnergyResolved}(b) shows how such highly entangled states emerge with increasing the disorder strength $\Delta$ with another $S^z_\mathrm{tot}$.
Although states $\ket{\mathcal{S}_k}$ have a sub-volume-law EE \cite{Shibata2020}, the disorder-averaged maximum entropies  $[S_\mathrm{vN}^\text{max}]$ exhibit a volume-law behavior, as shown in Fig.~\ref{Fig:EnergyResolved}(c).
Moreover, at large $\Delta$, Fig.~\ref{Fig:EnergyResolved}(d) shows that the ratio between the number $N$ of high-entanglement states and the Hilbert space dimension ${D}$ approaches zero in large system limit, i.e., $\lim _{L \rightarrow \infty} {[N]}/{D} \rightarrow 0$, signaturing that the inverted QMBS are also measure zero states, which shares the same spirit of conventional QMBS.
In the bulk $S_\mathrm{tot}^z$ sectors without states $\ket{\mathcal{S}_k}$, we also find
anomaly high entanglement states, and they concentrate in the middle of the energies of the exact eigenstates $\ket{\mathcal{S}_k}$. Therefore, the Hamiltonian \eqref{eq:totH} realizes multiple inverted QMBS concentrating in different narrow energy windows with approximately equal energy spacing $\approx 1$, which is the half of the energy spacing of states $\ket{\mathcal{S}_k}$ \cite{SM}. We remark that the number of high entanglement states in every energy window is much larger than one (as detailed in the caption of Fig.~\ref{Fig:EnergyResolved}(a)), and the narrow energy windows with highly entangled states also exhibit peaks  of the energy density of states (DOS) \cite{SM}.

We further understand the behavior of the inverted QMBS.
Indeed, we find these highly entangled states have a large overlap with the states $|{\phi_\alpha^{H_\mathrm{A}}}\rangle$ in the null space of $H_\mathrm{A}$ [see  Fig.~\ref{Fig:EnergyResolved}(e)], where $H_{A}|{\phi_\alpha^{H_\mathrm{A}}}\rangle=0$. As a result, such states stay delocalized and remain largely unaffected by the disorder strength.
Therefore, in both conventional and inverted QMBS, the Hamiltonian term $H_\mathrm{A}$ plays a pivotal role in dictating the thermalization properties of the total Hamiltonian. Specifically, it acts differently on these two kinds of measure zero states compared to other majority states.
However, while $\ket{\mathcal{S}_k}$ is written by a certain superposition of null vectors $\ket{\phi_\alpha^{H_\mathrm{A}}}$, the superposition constituting the inverted QMBS can be random \cite{SM}.
{
Similar to the conventional QMBS, for inverted QMBS there are also distinct experimentally observed signatures in dynamics.
Quenching from a random thermal state $|\psi(0)\rangle$ restricted in the null space of $H_{\mathrm{A}}$, the fidelity revives periodically in dynamics, as demonstrated in Fig.~\ref{Fig:EnergyResolved}(f).}

\section {Stability to onsite random field} Now we consider the stability of the aforementioned  exact QMBS $\ket{\mathcal{S}_k}$ and inverted QMBS to the onsite random $z$ fields that break the formalism $H$ \eqref{eq:framework}.
To be more specific, we modify $H$ in \eqref{eq:totH} to be
$
  H'=H+h \sum_{j=1}^{L} \delta_jS_{j}^{z},
$
where $\delta_j$ are the uniform random numbers in the range $\delta_j\in[-1,1]$. Unlike the disordered $H_\mathrm{A}$, the disorder term in $H'$ can drive all  eigenstates to the localization. In Figs.~\ref{Fig:c1h}, with a localized spectrum background, both the periodic revival of the fidelity for the initial state $\ket{\psi_0}$ and the peak of $[\overline{S_\mathrm{vN}/S_\mathrm{Page}}]$ show certain stability of both exact tower of eigenstates $\ket{\mathcal{S}_k}$ and inverted QMBS  against the onsite random $z$ field, though they eventually disappear for a large disorder strength $h$. In a thermalizing background, with the increase of $h$, the conventional QMBS $\ket{\mathcal{S}_k}$ first disappears and the system becomes thermal before the final localization of all the states  \cite{SM}, consistent with previous scenarios of the QMBS in the disordered PXP models \cite{KeHuang2021,MondragonShem2021}.

\section {Summary and outlook} In this work, we extend the framework of QMBS to construct multiple inverted QMBS states using random superpositions realized by a disordered version of
$H_\mathrm{A}$ with large amplitude and considerable null vectors.
With this generalized framework, we have realized a direct evolution from a thermal spectrum background with a tower of exact QMBS to the MBL background with multiple inverted QMBS in a disordered spin-1/2 XX Heisenberg chain. The forms of the exact tower of states are independent of the disorder, while the energy-level statistics of the bulk eigenstates changes from Wigner-Dyson to Poisson when increasing the disorder, despite the existence of the embedded highly entangled states at strong disorder.
Embedded in the otherwise MBL spectra with low entanglement, the multiple sets of many highly entangled states are located within different narrow energy windows that are approximately equidistant in energy.
To the best of our knowledge, such a scenario that inverts multiple QMBS directly is not constructed before.
Our model can also be generalized to other non-Abelian symmetry, and to the large classes of QMBS Hamiltonian that resort to the annihilating term $H_\mathrm{A}$. {The proposal to invert QMBS in this work may also serve as a catalyst for experimentalists to either develop new methods or adapt existing technique to realize setups that weakly violate the MBL \cite{Marcuzzi2017,Ostmann2019,VanVoorden2021,XXLi2022}}.

\emph {Note added.}  When finalizing the manuscript, we became aware of one  recent work \cite{Iversen2023} on related topics.

\begin{acknowledgments}

We are grateful to  Yi-Zhuang You, Shuai A. Chen, Zlatko Papi\'{c}, Jean Yves Desaules, Andrew Hallam, Yang Qi for the fruitful discussions.
This work was supported by the National Natural Science Foundation of China (Grant No.12074375), the Fundamental Research Funds for the Central Universities and the Strategic Priority Research Program of CAS (Grant No.XDB33000000).
\end{acknowledgments}

\begin{appendix}

\setcounter{equation}{0}
\setcounter{subsection}{0}
\renewcommand{\theequation}{\thesection.\arabic{equation}}

\section{Multiple inverted QMBS in different symmetry sectors.}

\begin{figure}[b]
\begin{center}
\includegraphics[width=0.48\textwidth]{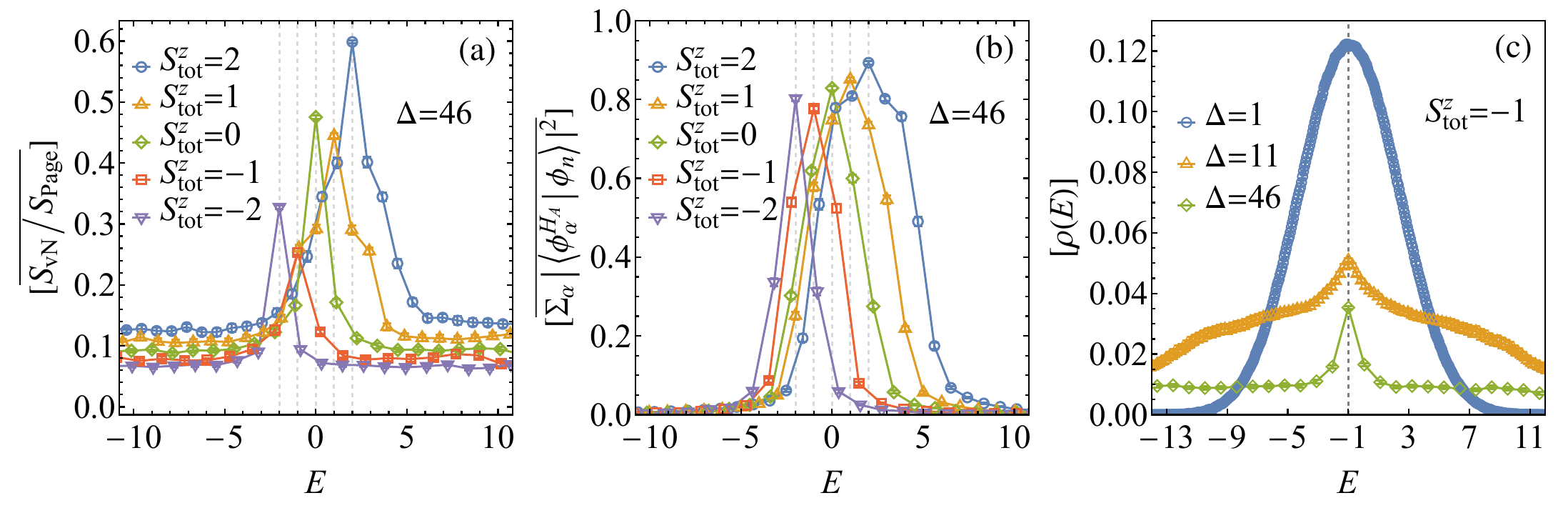}
\end{center}
\par
\renewcommand{\figurename}{Fig.}
\caption{ Features of multiple inverted QMBS in different $S_\mathrm{tot}^z$ sectors.
(a) The energy-resolved $[\overline{S_\text{vN}/S_\text{Page}}]$  for $L=16$.
(b) The overlap between eigenstates of $H$ (i.e.,$\ket{\phi_n}$) and   $\vert\phi_\alpha^{H_\mathrm{A}}\rangle$ for $L=16$, where $H_\mathrm{A}\vert\phi_\alpha^{H_\mathrm{A}}\rangle=0$.
(c) Density of states for $L=18$, $S^z_\text{tot}=-1$. At $\Delta=46$, the peak of DOS appears in the energy window of inverted QMBS. For comparison, the DOS for smaller  $\Delta$  are also plotted.
 }
\label{FigS1:MultipleSector}
\end{figure}

At strong disorder, multiple sets of highly entangled states concentrating in equidistant energy windows emerge in different $S_\mathrm{tot}^z$ sectors, as shown by the peaks of the energy-resolved $[\overline{S_\text{vN}/S_\text{Page}}]$ in Fig.~\ref{FigS1:MultipleSector}(a). We remark that these energy windows with peaks of $[\overline{S_\text{vN}/S_\text{Page}}]$ are indeed very narrow compared to the large width of the whole energy spectrum. For example, for $S_\mathrm{tot}^z=2$ sector of $L=16$ in Fig.~\ref{FigS1:MultipleSector}(a), the width of the whole energy spectrum is $\sim 336$, while the peak of $[\overline{S_\text{vN}/S_\text{Page}}]$ is only $\sim 8$. The spacing between these energy windows is roughly 1. Figure \ref{FigS1:MultipleSector}(b) shows that the highly entangled states are indeed almost annihilated by the term $H_\mathrm{A}$ and thus remain largely undisturbed by the disorder. Moreover, at large $\Delta$, we also find the peak of the averaged  density of states $[\rho(E)]$ appears at the narrow energy window where the highly entangled states locate, as shown by the typical sector $S_\mathrm{tot}^z=-1$ of $L=18$ in Fig.~\ref{FigS1:MultipleSector}(c).

\section{Understanding inverted QMBS from the null space of the annihilating term}

\begin{figure}[t]
\begin{center}
\includegraphics[width=0.48\textwidth]{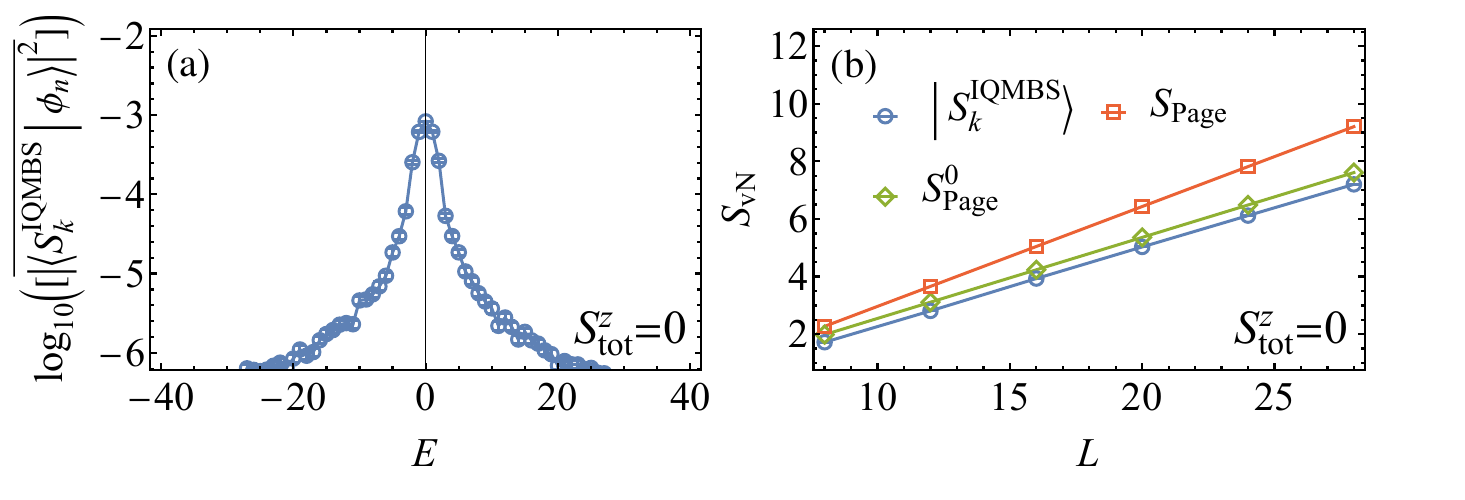}
\end{center}
\par
\renewcommand{\figurename}{Fig.}
\caption{
(a) The overlap between $|\mathcal{S}_k^{\mathrm{IQMBS}}\rangle$ and all the eigenstates. Average over states and disorder in each energy window has been applied. Here, $L=16$, $S_\mathrm{tot} ^z=0$, $\Delta=46$. (b) Entanglement entropy of $|\mathcal{S}_k^{\mathrm{IQMBS}}\rangle$ (blue circles) as a function of system size $L$.  $S_{\text {Page }}$ and $S_{\text {Page }}^0$ are also plotted for comparison, where $S_{\text {Page }}$ (red squares) is the Page value for a random pure state in the $S_{\text {tot }}^z=0$ sector, $S_{\text {Page }}^0$ (green diamonds) is the Page value in the null space of $H_{\mathrm{A}}$.
}
\label{Fig:IQMBSAnsatz4}
\end{figure}

The definition and the properties of inverted QMBS share a similar spirit to those of conventional QMBS. For both the inverted and conventional QMBS in our case, the annihilating Hamiltonian term that plays the key role in the thermalization properties of the total Hamiltonian acts completely differently on these two kinds of measure zero states compared to other majority states. In the Hamiltonian \eqref{eq:totH} in the main text, the disorder term $H_\mathrm{A}$ treats states $\prod_{j}(\mathbb{I}-\ketbra{010}{010}_{j-1,j,j+1}){\ket{n}}$ all the same, i.e., those states feel no disorder strength. Here, $\ket{n}$ is the product state. Since the highly entangled states live almost in such a disorder-free subspace spanned by $\prod_{j}(\mathbb{I}-\ketbra{010}{010}_{j-1,j,j+1}){\ket{n}}$, the effective disorder they feel is insufficient to cause their localization, unlike most other states experiencing a large disorder strength. For the conventional QMBS $\ket{\mathcal{S}_k}$, this situation is even more thorough, since conventional QMBS are completely annihilated by the disordered term in the Hamiltonian and thus feel no disorder at all. Notably, using the annihilating term to construct conventional QMBS is already widely used in various kinds of models and formalisms. However, while both inverted and conventional QMBS are largely superposed by the null vectors of the annihilating term, the ways of superposition are different, which makes the essential difference between the two.
To gain more understanding of the inverted QMBS, we propose a tentative wavefunction $|\mathcal{S}_k^{\text {IQMBS }}\rangle$ with random coefficients that can well capture the key signature of the inverted QMBS, analogous to the role played by the $|\mathbb{Z}_2\rangle$ state in the PXP scar model. Such a wavefunction can be written as
$$ |\mathcal{S}_k^{\mathrm{IQMBS}}\rangle = (1/\mathcal{N})
\sum_n \delta_n \prod_j(\mathbb{I}-|010\rangle\langle 010|_{j-1, j, j+1})|n_k\rangle,$$
where $\sum_jS_j^z\ket{n_k}=k$, $\mathcal{N}$ is the renormalization factor, and $\delta_\alpha$ to be the Gaussian random number. We will explain more about our considerations for choosing this wavefunction in the following.
We have also noticed that nearly every $$|\phi_n^{H_{\mathrm{A}}}\rangle\equiv \prod_j(\mathbb{I}-|010\rangle\langle 010|_{j-1, j, j+1})|n\rangle,$$
instead of just a few specific ones, contributes to the inverted QMBS eigenstates, and their contributions change dramatically from one disorder realization to another.
Thus, we choose $ |\mathcal{S}_k^{\mathrm{IQMBS}}\rangle$ to be superposed by all the product states $\left|\phi_n^{H_{\mathrm{A}}}\right\rangle $.
Moreover, the inverted QMBS states still experience a very small effective disorder strength, thus showing randomness and disorder-dependence. Since the eigenstate coefficients of a nonintegrable many-body Hamiltonian generally follow a random matrix theory prediction and display Gaussian distribution, we choose the coefficients $\delta_\alpha$ to be Gaussian random.
To confirm $|\mathcal{S}_k^{\text {IQMBS }}\rangle$ captures the key features of inverted QMBS, we confirm the inverted QMBS have a large overlap with the proposed $|\mathcal{S}_k^{\text {IQMBS }}\rangle$ compared with other eigenstates, as shown in Fig.~\ref{Fig:IQMBSAnsatz4}(a) for a representative $S_{\text {tot }}^Z=0$ sector. Furthermore, we notice that the largest value of $|\langle\mathcal{S}^{\text {IQMBS }} \mid \phi_n\rangle|^2$, as depicted in Fig.~\ref{Fig:IQMBSAnsatz4}(a), is comparable to the averaged overlap of the inverted QMBS states from different disorder realizations with identical parameters. This consistency is in line with the characteristic of a Gaussian random superposition for $|\mathcal{S}_k^{\text {IQMBS }}\rangle$. We also confirmed that the entanglement entropies of $|\mathcal{S}_k^{\text {IQMBS }}\rangle$ show a volume law scaling (see the blue circles in Fig.~\ref{Fig:IQMBSAnsatz4}(b)). Such behavior is very close to the Page entropy $S_{\text {Page }}^0$ (green diamonds) of the null space of $H_{\mathrm{A}}$ and the Page entropy $S_{\text {page }}$ (red squares) of $S_{\text {tot }}^z$ sectors. Thus, such a putative wavefunction captures the major features of inverted QMBS.

\begin{figure}[t]
\begin{center}
\includegraphics[width=0.45\textwidth]{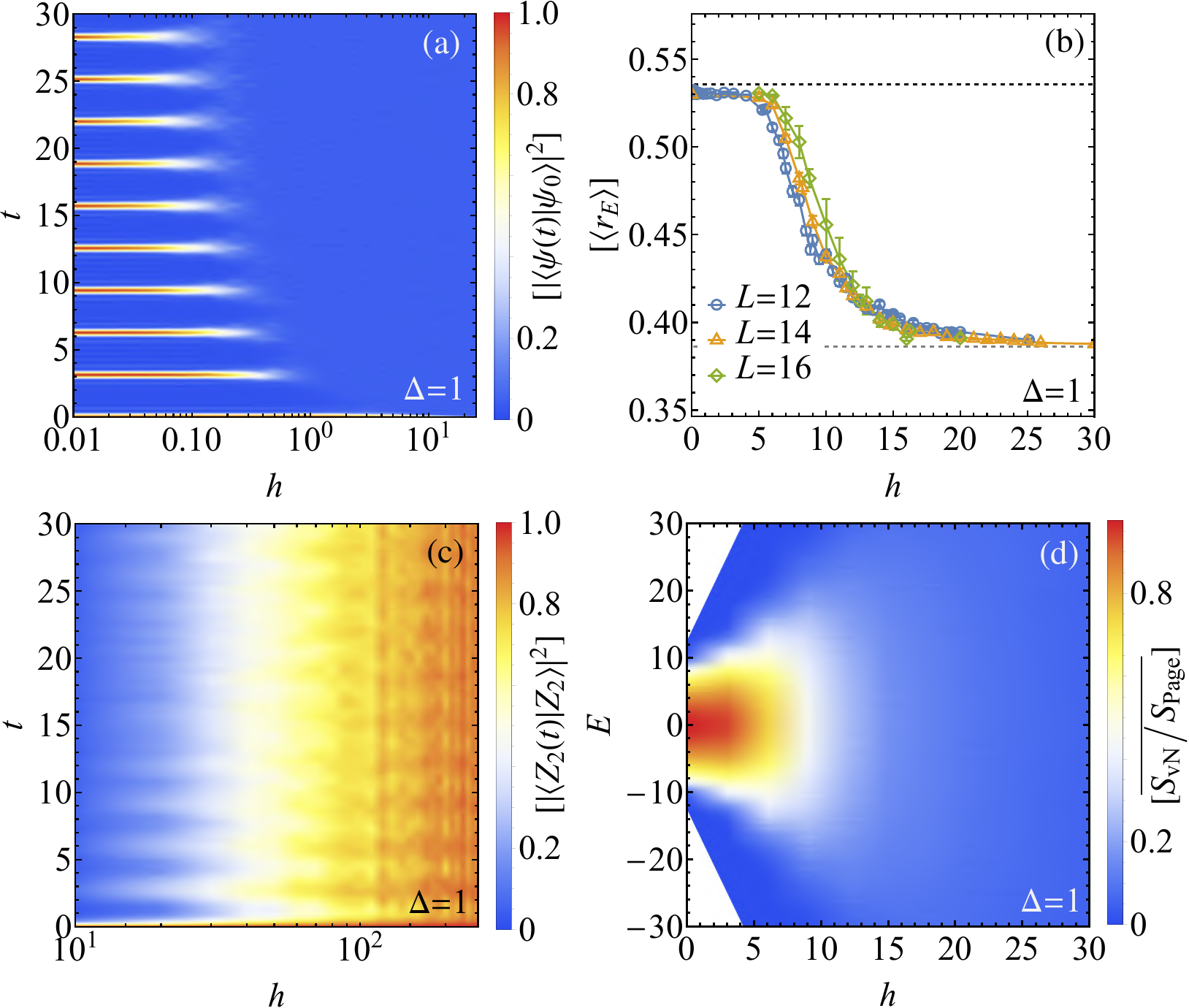}
\end{center}
\par
\renewcommand{\figurename}{Fig.}
\caption{Stability of QMBS $\ket{\mathcal{S}_k}$ at weak disorder strength $\Delta$.
(a) The disorder-averaged fidelity dynamics $[f(t)]=[|\braket{\psi(t)}{\psi(0)}|^{2}]$
of the initial  state $\ket{\psi(0)}\equiv\ket{\psi_0}$ (defined in the main text) with $L=12$.
(b) Mean level spacing ratios $\left[\left\langle r_E\right\rangle\right]$ as a function of $h$.
As a comparison, Wigner-Dyson  statistics of the GOE $\left\langle r_E\right\rangle\approx 0.536$ (dashed black lines) and Poisson  statistics $\left\langle r_E\right\rangle\approx 0.38$ (dashed gray lines) are plotted.
The $\left[\left\langle r_E\right\rangle\right]$ are averaged over 100 disorder realizations for $L=12,14$ and between 10 and 40 for $L=16$.
(c) The disorder-averaged fidelity dynamics $f(t)=|\braket{Z_2(t)}{Z_2}|^{2}$ of the initial  state $\ket{Z_2}$ with $L=14$ at different disorder strength $h$.
(d) The energy-resolved $[\overline{S_\text{vN}/S_\text{Page}}]$ as a function of $h$ for $L=12$.
 }
\label{Fig1:c123hDelta1}
\end{figure}
\section{Fate of QMBS with onsite random field}

In this section, we study the fate of QMBS in the presence of onsite random $z$ field $h$.
Here we consider a different annihilating disorder with more  terms
\begin{equation*}\label{}
\begin{split}
  H'_{A}=\Delta\sum_{j}\{
  &c^{(1)}_ { j }  | 0 1 0 \rangle \langle010|\\
  &+\frac{c_{j}^{(2)}}{2}(|011\rangle+|110\rangle)(\langle 011|+\langle 110|)\\
  &+c_{j}^{(3)}[|010\rangle(\langle 011|+\langle 110|)+\text {h.c.}]\}_{j-1, j, j+1},
\end{split}
\end{equation*}
where $c _ { j }^{(\alpha)}$ with $\alpha=1,2,3$ are  the uniform random numbers $c _ { j } ^ { ( \alpha ) }\in[-1,1]$. We remark that $H'_{A}$ breaks U(1) symmetry. The total Hamiltonian reads
\begin{equation}\label{eqS:totH}
\begin{aligned}
  H=&\sum_{j=1}^{L} \left(S_{j}^{+} S_{j+1}^{-}+S_{j}^{-} S_{j+1}^{+}+ S_{j}^{z}\right)+ H'_\mathrm{A}+h \sum_{j=1}^{L} \delta_jS_{j}^{z}.
\end{aligned}
\end{equation}
The last term in \eqref{eqS:totH} can be regarded as the onsite random fields that break the symmetry-based formalism mentioned in the main text.
Some characteristic features of QMBS, such as the slow relaxation from certain initial states, are still existent in the presence of a modest disorder strength $h$, as shown by Fig.~\ref{Fig1:c123hDelta1}(a). As $h$ is increased, however, the model \eqref{eqS:totH}  loses  the  QMBS features before switching  to MBL (c.f. Figs.~\ref{Fig1:c123hDelta1}(b-d)). Remarkably, in the MBL spectrum background, there is no peak of $[\overline{S_\text{vN}/S_\text{Page}}]$, since the random $z$ fields can affect every eigenstate.

\end{appendix}

\end{document}